\documentstyle[prl,aps]{revtex}

\begin{document}


\begin{center}
{\large\bf A Search for Jet Handedness in Hadronic $Z^0$ Decays}
\end{center}

\vspace{1 cm}

\begin{center}
%
%
%
  \def\iADEL{$^{(1)}$}
  \def\iBOL{$^{(2)}$}
  \def\iBU{$^{(3)}$}
  \def\iBRUN{$^{(4)}$}
  \def\iCIT{$^{(5)}$}
  \def\iUCSB{$^{(6)}$}
  \def\iUCSC{$^{(7)}$}
  \def\iCIN{$^{(8)}$}
  \def\iCSU{$^{(9)}$}
  \def\iCOLO{$^{(10)}$}
  \def\iCOL{$^{(11)}$}
  \def\iFER{$^{(12)}$}
  \def\iFRA{$^{(13)}$}
  \def\iILL{$^{(14)}$}
  \def\iLBL{$^{(15)}$}
  \def\iMIT{$^{(16)}$}
  \def\iMASS{$^{(17)}$}
  \def\iMISS{$^{(18)}$}
  \def\iNAG{$^{(19)}$}
  \def\iOREG{$^{(20)}$}
  \def\iPAD{$^{(21)}$}
  \def\iPERU{$^{(22)}$}
  \def\iPISA{$^{(23)}$}
  \def\iRUT{$^{(24)}$}
  \def\iRAL{$^{(25)}$}
  \def\iSOGANG{$^{(26)}$}
  \def\iSLAC{$^{(27)}$}
  \def\iTENN{$^{(28)}$}
  \def\iTOH{$^{(29)}$}
  \def\iVAND{$^{(30)}$}
  \def\iWASH{$^{(31)}$}
  \def\iWISC{$^{(32)}$}
  \def\iYALE{$^{(33)}$}
  \def\dead{$^{\dag}$}
  \def\andgen{$^{(a)}$}
  \def\andper{$^{(b)}$}

\mbox{K. Abe                 \unskip,\iTOH}
\mbox{I. Abt                 \unskip,\iILL}
\mbox{C.J. Ahn               \unskip,\iSOGANG}
\mbox{T. Akagi               \unskip,\iSLAC}
\mbox{W.W. Ash               \unskip,\iSLAC$^\dagger$}
\mbox{D. Aston               \unskip,\iSLAC}
\mbox{N. Bacchetta           \unskip,\iPAD}
\mbox{K.G. Baird             \unskip,\iRUT}
\mbox{C. Baltay              \unskip,\iYALE}
\mbox{H.R. Band              \unskip,\iWISC
% ...
}
\it
  \vskip \baselineskip                   
  \vskip \baselineskip                   
%
%
%
  \iADEL
     Adelphi University,
     Garden City, New York 11530 \break
  \iBOL
     INFN Sezione di Bologna,
     I-40126 Bologna, Italy \break
  \iBU
     Boston University,
     Boston, Massachusetts 02215 \break
  \iBRUN
     Brunel University,
     Uxbridge, Middlesex UB8 3PH, United Kingdom \break
  \iCIT
     California Institute of Technology,
     Pasadena, California 91125 \break
  \iUCSB
     University of California at Santa Barbara,
     Santa Barbara, California 93106 \break
  \iUCSC
     University of California at Santa Cruz,
     Santa Cruz, California 95064 \break
  \iCIN
     University of Cincinnati,
     Cincinnati, Ohio 45221 \break
  \iCSU
     Colorado State University,
     Fort Collins, Colorado 80523 \break
  \iCOLO
     University of Colorado,
     Boulder, Colorado 80309 \break
  \iCOL
     Columbia University,
     New York, New York 10027 \break
  \iFER
     INFN Sezione di Ferrara and Universit\`a di Ferrara,
     I-44100 Ferrara, Italy \break
  \iFRA
     INFN  Lab. Nazionali di Frascati,
     I-00044 Frascati, Italy \break
  \iILL
     University of Illinois,
     Urbana, Illinois 61801 \break
  \iLBL
     Lawrence Berkeley Laboratory, University of California,
     Berkeley, California 94720 \break
  \iMIT
     Massachusetts Institute of Technology,
     Cambridge, Massachusetts 02139 \break
  \iMASS
     University of Massachusetts,
     Amherst, Massachusetts 01003 \break
  \iMISS
     University of Mississippi,
     University, Mississippi  38677 \break
  \iNAG
     Nagoya University,
     Chikusa-ku, Nagoya 464 Japan  \break
  \iOREG
     University of Oregon,
     Eugene, Oregon 97403 \break
  \iPAD
     INFN Sezione di Padova and Universit\`a di Padova,
     I-35100 Padova, Italy \break
  \iPERU
     INFN Sezione di Perugia and Universit\`a di Perugia,
     I-06100 Perugia, Italy \break
  \iPISA
     INFN Sezione di Pisa and Universit\`a di Pisa,
     I-56100 Pisa, Italy \break
  \iRUT
     Rutgers University,
     Piscataway, New Jersey 08855 \break
  \iRAL
     Rutherford Appleton Laboratory,
     Chilton, Didcot, Oxon OX11 0QX United Kingdom \break
 \iSOGANG
     Sogang University, Seoul Korea \break
  \iSLAC
     Stanford Linear Accelerator Center, Stanford University,
     Stanford, California 94309 \break
  \iTENN
     University of Tennessee,
     Knoxville, Tennessee 37996 \break
  \iTOH
     Tohoku University,
     Sendai 980 Japan \break
  \iVAND
     Vanderbilt University,
     Nashville, Tennessee 37235 \break
  \iWASH
     University of Washington,
     Seattle, Washington 98195 \break
  \iWISC
     University of Wisconsin,
     Madison, Wisconsin 53706 \break
  \iYALE
     Yale University,
     New Haven, Connecticut 06511 \break
  \dead
     Deceased \break
  \andgen
     Also at the Universit\`a di Genova \break
  \andper
     Also at the Universit\`a di Perugia \break
\rm
%

\end{center}

\vspace{1 \baselineskip}

\begin{abstract}
We have searched for signatures of polarization in hadronic jets from
$Z^0 \rightarrow q \bar{q}$ decays using the ``jet handedness'' method.
The polar angle asymmetry induced by the high SLC electron-beam polarization
was used to separate quark jets from antiquark jets, expected to be left-
and right-polarized, respectively. We find no evidence for jet handedness
in our global sample or in a sample of light quark jets and we set upper
limits at the 95\% C.L. of 0.063 and 0.099 respectively on the magnitude
of the analyzing power of the method proposed by Efremov {\it et al.}
\end{abstract}

\pacs{ }

   The transport of parton polarization in strong interactions
is of fundamental interest.
It is at present an open
question whether the polarization of quarks or antiquarks
produced in hard collisions is
observable via the final-state fragmentation products in the resulting jets.
The $Z^0$ resonance is an ideal place to study this issue as the partons
in $Z^0 \rightarrow q\bar{q}$ decays
are predicted by the Standard Model (SM) to be highly longitudinally polarized.
If a method of observing such polarization were developed,
it could be applied to jets produced in a variety of hard processes in order to
elucidate the spin dynamics of the underlying interactions.

In the process $e^+e^- \rightarrow Z^0 \rightarrow q\bar{q}$ the relative
cross sections for production of left- and right-handed quarks of flavor $f$
are
given at tree level by \cite{sm}:
\begin{eqnarray}
\label{pq}
 \sigma^f_L & = & ( 1 + A_f )( 1 + \cos^2\theta + 2 A_Z \cos\theta )  \\
 \sigma^f_R & = & ( 1 - A_f )( 1 + \cos^2\theta - 2 A_Z \cos\theta ), \nonumber
\end{eqnarray}
where $A_Z = (A_e - P_{e^-})/(1 - A_e P_{e^-})$,
$A_f = 2 v_f a_f / (v_f^2 + a_f^2)$,
$P_{e^-}$ is the longitudinal polarization of the electron beam,
$v_f$ and $a_f$ are the vector and axial-vector couplings of fermion $f$ to the
$Z^0$,
and $\theta$ is the polar angle of the outgoing quark with
respect to the electron beam direction.
The quark and antiquark in a $Z^0$ decay have opposite helicities so that
$\sigma^{\bar{f}}_{L(R)}(\cos\theta) = \sigma^{f}_{R(L)}(-\cos\theta)$.
The SM predicts $A_{e,\mu,\tau} \approx 0.16$, $A_{u,c} \approx 0.67$ and
$A_{d,s,b} \approx 0.94$
so that the quarks are produced predominantly left-handed and
the antiquarks predominantly right-handed.
In order to observe a net polarization in an ensemble of jets from $Z^0$
decays it is therefore necessary to distinguish quark jets from antiquark jets.

This separation can be achieved
at the SLAC Linear Collider (SLC) where $Z^0$ bosons are produced in collisions
of highly longitudinally polarized electrons with unpolarized positrons.
In 1993 the average electron beam polarization was 0.630$\pm$0.011 \cite{alr}.
In this case the SM predicts a large
difference in polar angle distributions between quarks and antiquarks,
providing an unbiased separation of quark and antiquark jets.
We define the ``helicity-based'' polarization of jets at a given $\cos \theta$:
\begin{equation}
\label{ppol}
P_{hel}(\cos\theta)  \equiv
  \frac{ \sigma^f_R + \sigma^{\bar{f}}_R - \sigma^f_L - \sigma^{\bar{f}}_L }
       { \sigma^f_R + \sigma^{\bar{f}}_R + \sigma^f_L + \sigma^{\bar{f}}_L }
    = -2 \frac{A_Z \cos \theta }{ 1 + \cos^2 \theta} .
\end{equation}
This jet polarization is independent of flavor, and reaches 0.72 and 0.52 in
magnitude at large
$|\cos \theta|$ for beam polarizations of $-$0.63 and $+$0.63 respectively.
An alternative variable is the ``chirality-based'' polarization of jets:
\begin{equation}
\label{pchi}
P_{chi}^f  \equiv
    \frac{ \sigma^f_R - \sigma^{\bar{f}}_R - \sigma^f_L + \sigma^{\bar{f}}_L }
         { \sigma^f_R + \sigma^{\bar{f}}_R + \sigma^f_L + \sigma^{\bar{f}}_L }
      = -A_f.
\end{equation}
This jet polarization is independent of $\cos\theta$ and electron beam
polarization but depends on quark flavor.
It is accessible by charge ordering of the tracks used in the analysis as
described below.
The experimental challenge is to find observables sensitive to one or both of
these expected jet polarizations.

   Nachtmann \cite{nachtmann} and Efremov {\em et al}. \cite{efremov} have
speculated that the polarization may be observable
inclusively via a triple product of track momenta in jets.
Arguing that quark fragmentation may resemble a multi-body strong decay,
they note that the simplest
observable with the same transformation properties under parity inversion
as $P_{hel}$ (eqn. 2) has the form
$\Omega \equiv \vec{t} \cdot ( \vec{k}_1 \times \vec{k}_2 )$ ,
where $\vec{t}$ is a unit vector along the jet axis, corresponding to the spin
direction of a longitudinally polarized parton that produced the jet, and
$\vec{k}_1$ and
$\vec{k}_2$ are the momenta of two particles in the jet chosen by some
charge-independent prescription,
e.g.,
\begin{equation}
\label{hel}
\Omega_{hel} = \vec{t} \cdot ( \vec{k}_1 \times \vec{k}_2 )\quad \hbox{where}
\quad|\vec{k}_1| > |\vec{k}_2|,
\end{equation}
and referred to some suitable frame. Alternatively,
if $\vec{k}_1$ and $\vec{k}_2$ are the momenta of a positively and a negatively
charged particle, the cross product can be ordered by charge, e.g.,
\begin{equation}
\label{guess}
\Omega_{chi} = \vec{t} \cdot ( \vec{k}_+ \times \vec{k}_- ).
\end{equation}
For a given flavor $\Omega_{chi}$ has the same transformation properties under
parity inversion as $P_{chi}$ (Eq.~3) and so might be sensitive to $P_{chi}$.
Jets from $Z^0$ decays comprise a mixture of flavors that might yield
different signals since quark charges and fragmentation properties depend on
flavor.
Taking into account only the sign $s_f$ of the charge of quarks of flavor $f$,
one  expects a net polarization
$P_{chi} = -\Sigma_f s_f R_f A_f = 3R_d A_d - 2R_u A_u \approx 0.39$,
where $R_u \approx $0.17 and $R_d \approx $0.22 are the
SM fractions of hadronic $Z^0$ decays into $u\bar{u}$ or $c\bar{c}$ and
$d\bar{d}$, $s\bar{s}$ or $b\bar{b}$ respectively.

Although no quantitative estimate of the size of $\Omega$ is given,
it is argued in \cite{tornqvist} that such a term can arise from interference
between two processes, for example fragmentation into $\pi \pi \pi$ and
fragmentation into $\rho \pi$ where $\rho \rightarrow \pi \pi$.
Thus $\Omega$
might be largest for triplets of pions nearby in rapidity in which a
zero-charge pair has invariant mass near the $\rho$ mass, and the
unpaired track together with the oppositely charged track in the pair is
used to calculate $\Omega$ in the 3-pion rest frame.
It is also argued that evidence of polarization
is more likely to be visible in the highest-momentum tracks in jets.
Ryskin has proposed \cite{ryskin} an alternative physical
model of the transport of parton
polarization in the context of a string fragmentation scheme,
which gives a nonzero expectation value of
$\Omega$ in the laboratory frame if $\vec{k}_1$ and $\vec{k}_2$
are the momenta of two hadrons containing partons from
the same string breakup.

   A jet may be defined as left- or right-handed if $\Omega$ is negative
or positive respectively. For an ensemble of jets the jet handedness $H$
is defined as the asymmetry in the number of left- and right-handed jets:
\begin{equation}
\label{hdef}
  H \equiv \frac{N_{\Omega < 0} - N_{\Omega > 0}}
                {N_{\Omega < 0} + N_{\Omega > 0}}.
\end{equation}
It can then be asserted that
\begin{equation}
\label{hap}
  H = \alpha P,
\end{equation}
where $P$ is the expected
polarization of the underlying partons in the ensemble of jets, and $\alpha$
is the analyzing power of the method.

In this letter we present the results of the first search for jet handedness
in $Z^0 \rightarrow q \bar{q}$ decays using a
sample of approximately 50,000 hadronic $Z^0$ decays
collected by the SLD experiment \cite{sld} in 1993.
We have applied
the methods suggested in \cite{efremov,tornqvist} and \cite{ryskin},
and have extended them to be more inclusive.
For each method we used both helicity- and
chirality-based definitions of $\Omega$ to calculate $H$.
A handedness signal may be diluted in heavy flavor events,
$Z^0 \rightarrow c\bar{c}$ or $b\bar{b}$,
since a large fraction of the tracks in each jet is from the decay of
a spinless heavy meson.
Dalitz {\it et al.} have concluded \cite{dalitz} that
any effect resulting from $D^*$ or $B^*$ decays should be very small.
We therefore divided our data into samples enriched in light,
$Z^0 \rightarrow u\bar{u}$, $d\bar{d}$ or $s\bar{s}$, and
heavy flavor events using hadron lifetime information,
and sought evidence for jet handedness in each.

The trigger and initial selection of hadronic events is described in
\cite{alr}.
The analysis presented here is based on charged tracks measured in the
central drift chamber and vertex detector.
A set of cuts was applied to select
events well-contained within the detector acceptance.
Tracks were required to have
(i) a closest approach to the beam axis within 5~cm and within
10 cm along the beam axis of the measured interaction point,
(ii) a polar angle $\theta$
with respect to the beam axis with $|\cos\theta|$ $<$ 0.80, and
(iii) a minimum momentum transverse to this axis of $p_{\perp}$ $>$
150 MeV/$c$.
Events were required to contain a minimum of five such tracks,
a thrust \cite{thrust} axis polar angle with respect to the beam axis
$\theta_T$ within $|\cos\theta_T|$ $<$ 0.71, and
a minimum charged visible energy $E_{vis} > 20$ GeV,
where all tracks were assigned the charged
pion mass.
Events with hard gluon radiation were rejected by requiring events to contain
only two jets, defined using the JADE clustering algorithm \cite{jade}
at $y_{cut} = 0.03$, which were back-to-back within $20^{\circ}$.
A data sample comprising 17,853 events passed these cuts.

In addition to considering this global sample, events were classified as being
of light or heavy flavor based on impact
parameters of charged tracks measured in the vertex detector.
The 9,977 events containing no track with normalized transverse
impact parameter with respect to the interaction point $b/\sigma_b>3$
were assigned to the light-flavor sample and all other events
were assigned to the heavy-flavor sample.
The light-flavor contents of these two samples were estimated from Monte Carlo
simulations to be 84\% and 30\% respectively \cite{mikeh}.

Following \cite{tornqvist}
we first considered jets in which the three highest-momentum tracks
had total charge $\pm$1 and the
invariant mass of each zero-charge pair satisfied
$0.6 < m < 1.6$ GeV/$c^2$.
All tracks were assigned the charged pion mass and their momenta were boosted
into the three-track rest frame.
The tracks forming the higher-mass zero-charge pair were used to calculate
$\Omega_{hel} = \hat{t} \cdot ( \vec{k}_1 \times \vec{k}_2 )$ and
$\Omega_{chi} = \hat{t} \cdot ( \vec{k}_+ \times \vec{k}_- )$, where
$|\vec{k}_1| > |\vec{k}_2|$
and $\hat{t}$ is the thrust axis signed so as to point along the jet direction.
A signal would be visible as a nonzero mean $\Omega$, which in the case of the
helicity-based analysis would be of opposite sign for events produced with
positive and negative electron-beam polarization and for jets with positive and
negative
$\cos\theta = \hat{t}_z$.
The distributions of sgn($P_{e^-} \cos\theta)\Omega_{hel}$ and
$\Omega_{chi}$ are shown for the light-flavor sample in Fig.~1.
Both distributions appear to be symmetric about zero, implying that any
jet handedness is small.
Also shown in Fig.~1 are the predictions of the JETSET \cite{jetset}
Monte Carlo simulation program for $Z^0 \rightarrow q\bar{q}$ decays,
in which spin transport was not simulated,
combined with a simulation of the SLD.
These simulations give a good description of our measured
inclusive track and event topology distributions \cite{ascomp}.
The means of the simulated $\Omega$ distributions are consistent with zero
within statistical errors, limiting any analysis bias to 0.008
on $H$.
The variances of the measured distributions are reproduced by the simulation to
within 5\% relative, although the details of the shapes are not.

The jet handedness for the helicity-based analysis was calculated in bins of
jet
$\cos\theta$ according to Eq.~(6) separately
for events produced with positive and negative electron-beam polarization.
Results are shown in Fig.~2 for the light-flavor sample;
similar results (not shown) are obtained for the global and heavy-flavor
samples.
In all cases the measured jet handeness is consistent with zero, and there is
no evidence for an angular dependence.
Equation~(7) was fitted simultaneously
to the $H(\cos\theta)$ measured in events produced with positive and negative
electron-beam polarizations,
by averaging $P=P_{hel}$ (Eq.~2) over each
$\cos\theta$ bin and allowing the analyzing power $\alpha$ to vary.
The result of the fit to the light-flavor sample
is shown as the solid lines in Fig.~2, and the fitted analyzing powers for all
three samples are listed in Table I.

The jet-handedness values for the chirality-based analysis were calculated from
the unbinned $\Omega_{chi}$ distributions and the analyzing
powers were calculated from Eq.~(7), where $P=P_{chi}$ (Eq.~3) was
averaged over the flavor composition of each sample, estimated from the
simulations and weighted by the sign of the quark charge.
The analyzing powers are shown in Table I.
Since all $\alpha$ are consistent with zero, we set
upper limits at the 95\% confidence level on their magnitudes,
also shown in Table I.

Since the helicities of the quark and antiquark in a given event are
opposite, one might expect a correlation between $\Omega$ values
in the two jets in an event, which would be negative for the helicity-based
analysis and positive for the chirality-based analysis.
We found correlation coefficients to be consistent with zero, within
statistical
errors of $\pm$0.02, for both analyses and for the
global, light-, and heavy-flavor samples.

We extended this method to use the $N_{lead}$ highest-momentum
particles in each jet, with $3 \leq N_{lead} \leq 24$.
We considered all zero charge pairs $i$,$j$
among these $N_{lead}$ particles, without imposing mass cuts,
and calculated $\Omega_{hel}^{ij}$ and $\Omega_{chi}^{ij}$
for each pair in the $N_{lead}$-particle rest frame.
Jets with fewer than $N_{lead}$ tracks were excluded.
We then considered both the average, $<\Omega^{ij}>$, and the
$\Omega^{ij}$ with largest magnitude $\Omega^{max}$.
In both cases the jet handedness calculated
from the global, light- and heavy-flavor samples was consistent with zero
for all $N_{lead}$ and for both helicity- and chirality-based analyses.
For $N_{lead} \leq 11$ upper limits on the magnitudes of the analyzing powers
in the range 0.05-0.11 were derived.
For larger $N_{lead}$ the reduced sample size limits our accuracy.

Following \cite{ryskin} we then attempted to select pairs of tracks
likely to contain partons from the same string breakup.
In studies using JETSET we found
the relative rapidity with respect to the jet axis of the tracks in a pair
to be useful for this.
Requiring zero charge does not improve this selection, but was used
in the chirality-based analysis.

In each jet the
tracks were ordered in rapidity and assigned a number
$1 \leq n_i \leq n_{tracks}$,
where $n_i=1$ for the track with highest rapidity.
We then required pairs of tracks $i$,$j$ to have $|n_i - n_j| < \Delta n$ and
max($n_i$,$n_j$)$\leq n_{max}$.  Since the signal is expected \cite{ryskin}
to increase with momentum $p_t$ transverse to the thrust axis, we also required
$|p_{ti}|+|p_{tj}|>p_{min}$.
We calculated $\Omega_{chi}^{ij}$ and $\Omega_{hel}^{ij}$
in the laboratory frame for each pair satisfying these criteria
and then considered both the average, $<\Omega^{ij}>$, and the
$\Omega^{ij}$ with largest magnitude $\Omega^{max}$.
We varied $\Delta n$, $n_{max}$, and $p_{min}$ in an attempt to
maximize the handedness signal.
In all cases the jet handedness calculated from
the global, light- and heavy-flavor samples was consistent with zero.
We obtained upper limits in the range 0.05-0.11
for $n_{max} \leq 6$, $\Delta n \leq 6$ and $p_{min} <$2 GeV/c.
Statistics become poor in the potentially interesting high-$p_{min}$ region.

A number of systematic checks was performed.
The results of all analysis methods were found to be insensitive to
the track and event selection cuts, to the jet-finding algorithm
and $y_{cut}$ values used to select 2-jet events, and to
the values of the selection criteria for tracks used to define $\Omega$.
Each analysis was performed on samples of Monte Carlo events
in which spin transport was not simulated, yielding $H$ consistent with zero
within $\pm 0.004$.

In conclusion,
we have searched for evidence of parton polarization in hadronic $Z^0$
decays using the jet handedness methods proposed in
\cite{efremov,tornqvist} and \cite{ryskin}.
In an attempt to optimize a signal we studied a wide range of parameters for
each method.
In each case we applied both helicity- and chirality-based analyses,
and sought signals separately
in samples of light- and heavy-flavor jets as well as in the global sample.
We found no evidence for a non-zero jet handedness,
implying that the transport of polarization through the
quark fragmentation process is small. The method proposed in \cite{tornqvist},
applied to a sample of light-flavor jets,
yielded upper limits of 0.099 and 0.070
on the magnitudes of the analyzing powers of helicity- and chirality-based
analyses respectively.
Similar limits were derived for all other methods we applied.

We thank the personnel of the SLAC accelerator department and the technical
staffs of our collaborating instutions for their outstanding efforts on our
behalf.
We thank R. Dalitz, L. Dixon, A. Efremov, R. Jaffe, M. Peskin and M.~Ryskin
for helpful comments relating to this analysis.

\clearpage

\section*{Figure Captions}

\begin{figure}
\caption{
Measured distributions of (a) sgn($P_{e^-} \cos\theta)\Omega_{hel}$ and
(b) $\Omega_{chi}$ (points with error bars) for the light-flavor sample.
The corresponding distributions from a Monte Carlo simulation are also
shown (histograms).}
\label{fig1}
\end{figure}

\begin{figure}
\caption{
Helicity-based jet handedness as a function of jet $\cos\theta$ for the
light-flavor sample using jets produced with (a) negative and (b) positive
electron-beam polarization. The solid lines are the result of a fit of
Eq.~(6).}
\label{fig2}
\end{figure}

\clearpage
\section*{Table Captions}
\begin{table}
\caption{Analyzing powers of the helicity- and chirality-based definitions of
jet handedness.
Upper limits at the 95\% C.L. on the magnitudes are shown in parentheses.}
\end{table}

\clearpage
\setcounter{table}{0}
\begin{table}[hbt]
\label{resultsb}
\begin{center}
\begin{tabular}{lrrr} \hline
         & \multicolumn{3}{c}{ }             \\[-.4cm]
Analysis & \multicolumn{3}{c}{Analyzing Power}   \\
         & \multicolumn{1}{c}{Light-flavor jets}
         & \multicolumn{1}{c}{Heavy-flavor Jets}
         & \multicolumn{1}{c}{All Jets}      \\[.1cm] \hline
&&&\\[-.4cm]
Helicity            &--0.051$\pm$0.029  (0.099) &  0.000$\pm$0.032  (0.063)
                    &  0.028$\pm$0.022  (0.063)  \\
Chirality           &--0.018$\pm$0.031  (0.070) &--0.033$\pm$0.043  (0.104)
                    &--0.024$\pm$0.026  (0.066)  \\ \hline
\end{tabular}
\caption{}
\end{center}
\end{table}

\end{document}